# Using Ensemble Analysis to study the effects of Network Topology on Performance in Urban Road Networks

Steven J O'Hare, Richard D Connors, David P Watling

Institute for Transport Studies, University of Leeds, Leeds, United Kingdom

Corresponding Author: R.D.Connors@its.leeds.ac.uk

## Abstract

In recent years, researchers in the field of Network Science have begun to study the structural properties and performance characteristics of urban road networks. Such studies of the effect of network structure on performance have, thus far, been restricted to examining ensembles of synthetic networks generated by canonical models from the Network Science literature, which do not plausibly represent real road networks. Furthermore they tend to use a disparate range of parameter settings for demand and supply structure, making it difficult to compare one study with another.

To address these deficiencies, this paper proposes an experimental framework for the design of numerical investigations to study how network performance varies with respect to the structural properties of urban road networks. The key step within this experimental framework is the generation of a spectrum of ensembles of synthetic networks, which vary with respect to a single aspect of network structure and which adhere to plausibility constraints that ensure relevance to real urban road networks. This paper then demonstrates the application of this experimental framework to an investigation of how two performance indicators; the average link Volume-to-Capacity ratio and the Price of Anarchy, vary with respect to the density of travel demand, and the size, density and connectivity of network infrastructure. In so doing, several key challenges of such investigations are identified, with respect to the computational burden of the experiments and the need for empirical data on some important aspects of network structure such as the distribution of capacity.

## 1 Introduction

Over the last twenty years, advances in instrumentation and computing have significantly increased the amount of data that exists on a wide range of natural and manmade systems. The increasing availability of data has stimulated the emergence of a new field of research, called Network Science, in which researchers, from a variety of disciplines, have begun to use these data in the study of networked systems. The aims of Network Science include: 1) to characterise the structural properties of networks that underlie real-world systems, 2) to develop models that explain the formation of such structures, and 3) to investigate how the structural properties of networks affect the emergent characteristics of the systems they support (Newman, 2010).

Within this trend, urban road networks have been one of the many subjects of study. There have been empirical studies seeking to describe the structural characteristics of road infrastructure and patterns of travel demand in urban areas from across the world (Barthelemy, 2011). Theoretical studies have proposed generative models of network growth, describing the formation and evolution of urban road networks over time (Barthelemy and Flammini, 2009, Courtat et al., 2011). The effect of network structure on the performance of transportation networks have also been investigated, using large ensembles of synthetically generated networks and traffic modelling techniques (Sun et al., 2012, Wu et al., 2008b, Wu et al., 2008a, Wu et al., 2006, Youn et al., 2008, Zhao and Gao, 2007, Zhu et al., 2014).

The Network Science approach to the study of networks is recognised as a new perspective from which urban areas can be studied and understood (Batty, 2008). However, as an emerging research field, Network Science is not without criticism, and there are still opportunities for researchers from other fields to make significant contributions towards its development (Alderson, 2008, Havlin et al., 2012). This is certainly the case for urban road networks. For example, whilst empirical studies have identified both similarities and differences between the topological and geometric structures of road networks (Barthelemy, 2011), it is not clear that they have captured the full range of structure that exist in urban areas across the world, or even whether the measures that have been used thus far, quantify properties that are useful (Newman, 2003). Proposed generative models of road network growth have predominantly focussed on the formation of an appropriate topological and geometric structure, and have omitted other important features of road infrastructure. Finally, studies of the influence of network structure on performance in transportation networks have, thus far, focussed on performance comparisons between synthetic networks, generated by canonical models from the Network Science literature, which do not provide plausible representations of real urban road networks. Moreover, these studies lack a systematic experimental approach, making it difficult to generalise their findings or apply them to other families of networks. There is therefore significant scope for further developments within each of these research themes.

Looking to the transportation literature, it is evident that the transportation research community has made contributions in these areas, particularly with respect to the third theme of the effects of road network structure on network performance. Recent examples include Parthasarathi and Levinson (2010) and Levinson (2012), who collected real data for fifty cities in the United States of America (USA), and used regression techniques to study correlations between measures of network structure and urban mobility indicators. Their main finding was that cities with larger populations are typically more congested and have longer journey to work travel times. Tsekeris and Geroliminis (2013) used the theory of Macroscopic Fundamental Diagrams for a concentric city model to show that an increase in compactness, as travel demand increases, maximises efficiency with respect to congestion. Finally, Ortigosa and Menendez (2014) used traffic equilibrium modelling techniques to study the effects of the removal of links from a grid network, finding that a strategy of link removals from the geometric centre of a network is the most detrimental to performance.

Given these contributions, it is reasonable to question what Network Science has to offer in comparison with existing approaches in the transportation literature. The key difference between the two research disciplines is that Network Science searches for broad-scale commonalties in network phenomena across a wide range of network structures, whereas approaches in transport are typically location specific and focus on individual case studies. Evidence for the former can be seen in empirical studies of the structural properties of road networks, which draw upon data from a large number of urban areas. This contrasts with the approaches of, for example, Tsekeris and Geroliminis (2013) and Ortigosa and Menendez (2014) who focus exclusively on a concentric city and grid network respectively. This narrow focus makes it difficult to evaluate the generalizability of their findings with respect to the wide array of structures that are known to exist in real urban road networks. The approach used by Parthasarathi and Levinson (2010) and Levinson (2012) is a notable exception here because it uses data for a large number of different urban areas to search for relationships between network structure and performance; thereby bridging the gap between the transportation literature and Network Science. Further refelections on this follow, later in this paper.

This paper follows the approach of using large ensembles of synthetically generated networks to study how network structure affects performance. Keeping in mind the areas for development identified above, the goal of this paper is to explore how this technique can be more appropriately applied to study the variation of performance with respect to structure in the specific context of urban road networks. The main contribution of this paper is an experimental framework for the design of numerical investigations towards this aim. This framework sets out a systematic approach for tackling what is a high dimensional problem, which addresses the deficiencies of existing approaches, and which is generally applicable to a wide range of research questions. As a second contribution, the paper then demonstrates the application of this framework to several specific research questions, which focus on how two performance indicators; average link Volume-to-Capacity ratio and the Price of Anarchy, vary with respect to the density of travel demand, and the size, density and connectivity of network supply structure. As part of this investigation, this paper presents a simple model of road network generation, which is characterised by a small number of parameters, and is also able to produce a spectrum of synthetic network ensembles, which provide plausible representations of urban road networks and which also vary with respect to each of the aforementioned structural dimensions. Several challenges and opportunities for further research are also identified throughout the paper; in particular, with respect to the computational burden of numerical experiments and the lack of empirical data on several key aspects of urban road network structure.

The remainder of the paper is structured as follows. Section 2 presents a review of the relevant contributions that Network Science has made to the study of urban road networks. This review focusses on empirical studies of the structural properties of urban road networks, and numerical studies of the effects of network structure on performance, which have used the aforementioned synthetic networks approach. Section 3 provides a discussion of the main challenge that numerical investigations must tackle and describes the proposed experimental framework. Section 4 demonstrates the application of this experimental framework to a numerical investigation of how the two performance indicators vary with respect to the four dimensions of network structure identified above. Section 4 also includes a description of the model of road network generation and associated parameter settings used, and presents the results of the experiments. The paper concludes in section 5 with the identification of areas for further research and a discussion of how the proposed framework could be further developed and utilised.

## 2 Contributions from Network Science to the study of Urban Road Networks

### 2.1 Empirical Studies of the Structural Properties of Urban Road Networks

Network Science studies of the structure of urban road networks have, thus far, primarily focussed on the structural properties of supply (the physical infrastructure of junctions and roads) rather than travel demand (the aggregate movements of individual travellers between locations in the network). This is reflected in sections 2.1.1 and 2.1.2, which address each of these areas in turn.

#### 2.1.1 The Structural Properties of Supply in Urban Road Networks

Empirical studies of supply structure typically focus on a sample of urban areas and are based upon network data from geographic databases. Examples of data sources include: the NXI GESTATIO laboratory database, which was used by Buhl et al. (2006) for forty-one urban areas in Africa, Asia, Europe and South America; the Tele Atlas MultiNet$^{TM}$ geographic database, which was used by Lammer et al. (2006) and Chan et al. (2011) for twenty urban areas in Germany; the TIGER database, which was used by Jiang (2007) and Zhang et al. (2011) for forty-one urban areas in the USA; and the Ordnance Survey and Integrated Transport Network (ITN) datasets, which were used by Masucci et al. (2009) and Gudmundsson and Mohajeri (2013) for forty-one urban areas in the United Kingdom.

Once extracted, the network data for each urban area are processed to remove any network that is outside the area of interest and to simplify the level of detail. The remaining network data are then converted into an undirected graph of nodes, which represent junctions, and edges (or links), which represent road segments between junctions. This graph is referred to as a *primal* representation of road network structure (Porta et al., 2006)[1]. The structural properties of these primal networks are characterised by measures that quantify the characteristics of individual network components; the nodes, links and cells[2]. The statistical distributions of such measures over all components within the network are also examined. Examples of measures include node degree, $k_j$, which counts the number of links connected to a node $j$; and meshedness ($M$), which quantifies the overall level of connectivity in a network and is calculated as $M = m - n + 1/2n - 5$, where $n$ is the number of nodes and $m$ is the number of links (Buhl et al., 2006). A comprehensive list of measures relevant to road networks can be found in Barthelemy (2011).

Using measures such as those listed above, empirical studies have found that, at the *microscopic* level of nodes, links and cells, supply networks in a large number of urban areas share several common structural features. Perhaps unsurprisingly, urban road networks are predominantly *planar* (Buhl et al., 2006); where planarity means that links intersect each other only at nodes. Planarity significantly restricts other structural features in road networks; such as the average node degree, $\langle k \rangle$, to values $\langle k \rangle \leq 6$ (Barthelemy, 2011). Indeed, in a study of the twenty largest cities in Germany, Chan et al. (2011) reported the very narrow range of $\langle k \rangle \in [3.17, 3.31]$; a finding consistent with that of Buhl et al. (2006). Turning to link based measures, another common structural feature in most urban areas is that the frequency distribution of the angles formed between links at nodes is peaked around 90° and 180°, as occurs in gridded structures (Chan et al., 2011, Strano et al., 2013). Chan et al. (2011) argued that this is due to the fact that perpendicular roads minimise link lengths and construction costs, that rectangular cells are more desirable for buildings and that 90° angles provide desirable turning radii for vehicles. On the radial distribution of nodes, links and cell areas, Masucci et al. (2009) and Chan et al. (2011) showed that London and cities in Germany have a high density of nodes, short links and small cell areas in their city centres, but that their networks become more dispersed as distance from the centre increases. Related to this feature, Lammer et al. (2006) and Masucci et al. (2009) found that the distribution of cell areas in the cities of Dresden and London can be characterised by power laws.

Overall, the above results suggest that there are several common rules that govern the structure of supply in urban road networks. However, at the *macroscopic* level, the Network Science literature also points to a high degree of structural variation between road networks in different urban areas, particularly with respect to the size, density and connectivity of network infrastructure. For example, Chan et al. (2011) showed that urban areas in Germany vary considerably with respect to the geographical extent and the number of nodes in their road networks; see Table 1. With respect to the density of network infrastructure, Chan et al. (2011) also illustrated that the number of nodes per square kilometre varies considerably, from 9.8 nodes per km² in Bielefeld to 35.6 nodes per km² in Munich. Finally, with respect to network connectivity, Courtat et al. (2011)

---

[1] There is also a *dual* representation, in which nodes represent streets and edges signify the intersection of two streets (Porta et al., 2006a). Studies that use the dual representation are omitted because this approach removes all geometric data, which is an important influence on traffic flow behaviour.
[2] The city blocks that remain when the network links and nodes are subtracted from the 2D plane, $\mathbb{R}^2$

found a range of values for the meshedness measure across ten French cities, with values $M \in [0.2, 0.47]$; see Table 2. Cardillo et al. (2006) and Buhl et al. (2006) also found considerable variability in meshedness.

The findings of these studies are relevant for several reasons. Firstly, the similarities in structure, which have been identified at the *microscopic* level, suggest that road networks across a wide range of urban areas share a range of common structural characteristics. In contrast, the differences in structure at the *macroscopic* level highlight aspects of network structure whose effects on network performance should be explored. These findings reveal, and quantify, the extent to which road network supply structures are different from each other, and thereby motivate questions about the effects of such differences on performance, while also indicating reasonable ranges for these descriptive statistics in real networks.

| City | Population | Area (km²) | Nodes ($n$) | Population Density (Pop/km²) | Node Density ($\rho_n$) |
|---|---|---|---|---|---|
| Berlin | 3,392,425 | 891 | 19,931 | 3,807 | 22.4 |
| Hamburg | 1,728,806 | 753 | 9,044 | 2,296 | 12 |
| Munich | 1,234,692 | 311 | 11,058 | 3,970 | 35.6 |
| Cologne | 968,639 | 405 | 5,395 | 2,392 | 13.3 |
| Frankfurt | 643,726 | 249 | 3,911 | 2,585 | 15.7 |
| Dortmund | 590,831 | 281 | 3,281 | 2,103 | 11.7 |
| Stuttgart | 588,477 | 208 | 3,612 | 2,829 | 17.4 |
| Essen | 585,481 | 210 | 4,093 | 2,788 | 19.5 |
| Dusseldorf | 571,886 | 218 | 3,124 | 2,623 | 14.3 |
| Bremen | 542,987 | 318 | 3,827 | 1,708 | 12 |
| Duisburg | 508,664 | 233 | 2,837 | 2,183 | 12.2 |
| Leipzig | 494,795 | 293 | 3,753 | 1,689 | 12.8 |
| Nuremberg | 493,397 | 187 | 3,543 | 2,638 | 18.9 |
| Dresden | 480,228 | 328 | 3,346 | 1,464 | 10.2 |
| Bochum | 388,869 | 146 | 2,233 | 2,663 | 15.3 |
| Wuppertal | 363,522 | 168 | 1,750 | 2,164 | 10.4 |
| Bielefeld | 324,815 | 259 | 2,546 | 1,254 | 9.8 |
| Bonn | 308,921 | 141 | 2,094 | 2,191 | 14.9 |
| Mannheim | 308,759 | 145 | 2,674 | 2,129 | 18.4 |
| Karlsruhe | 281,334 | 173 | 2,204 | 1,626 | 12.7 |
| *Minimum* | *281,334* | *141* | *1,750* | *1,254* | *9.8* |
| *Average* | *740,063* | *296* | *4,713* | *2,355* | *15.5* |
| *Maximum* | *3,392,425* | *891* | *19,931* | *3,970* | *35.6* |

Table 1 - Network Size and Density of Supply Networks in twenty German Cities (Chan et al., 2011)

| City | Connectivity ($M$) | City | Connectivity ($M$) |
|---|---|---|---|
| Angoulme | 0.28 | Grenoble | 0.32 |
| Avignon | 0.23 | Lyon | 0.47 |
| Caen | 0.29 | Rennes | 0.26 |
| Carcassonne | 0.2 | Rouen | 0.38 |
| Dijon | 0.33 | Troyes | 0.28 |

Table 2 - Network Connectivity of Supply Networks in ten French Cities (Courtat et al., 2011)

Although these results display both similarities and differences between the road networks of different urban areas, they should be viewed with the caveat that the extent of their generality is unclear for the following three reasons. Firstly, there is significant variation between studies with respect to how raw network data are processed. For example, in setting the boundary for each urban area, Jiang (2007) used administrative regions, whereas Masucci et al. (2009) used a circle, centred on the city centre. Furthermore, in simplifying network data, Chan et al. (2011) chose to remove all nodes representing dead-ends, whereas Masucci et al. (2009) chose to retain them. These inconsistencies make it difficult to compare results between different studies. Secondly, all empirical studies, published to date, omitted data on junction types and other characteristics of network links, such as their capacity or position in the road hierarchy. These structural features influence traffic flow but their structural characteristics are not understood. Thirdly, and finally, each study used a different selection of network measures to characterise structure in a unique sample of cities, with few cases where the *same* network measure was applied to different datasets. With such limited crossover between studies, it cannot be concluded that all urban road networks share all of the properties described in this section. There is therefore a need for further empirical studies of the structural properties of urban road networks, which should aim to address the methodological inconsistencies, using an approach that focuses on the analysis of structure with traffic flow applications in mind, and which uses a broad selection of network measures to study a large sample of urban areas.

### 2.1.2 The Structural Properties of Demand in Urban Road Networks

In comparison with studies of the structure of network supply, empirical studies of the structure of travel demand are more limited in both quantity and the level of detail. This is primarily a reflection of the difficulty of collecting demand data; for which contributory factors include the greater fluidity in travel demand, which changes in both structure and magnitude on both an hourly and daily basis; the greater expense of common survey methods, such as roadside interviews and travel diaries; and the fact that such data are often based on responses from members of the public, which are prone to bias and typically achieve only low sample rates.

Of the studies that do exist, not one focuses exclusively on the pattern of travel demand in urban road networks. The focus has instead been on broader spatial scales or on the distribution of demand for other travel modes where data are more easily accessible. Examples include De Montis et al. (2007), who focussed on the distribution of travel demand between municipalities on the island of Sardinia; Patuelli et al. (2007), who focussed on the distribution of travel demand between major cities in Germany; and Jung et al. (2008), who focussed on the distribution of travel demand between major cities in South Korea. In each case, the distribution of demand was found to be highly heterogeneous, with the existence of few high volume movements and a large number of low volume movements. Heterogeneous distributions of demand have also been identified in urban areas by Chowell et al. (2003), who used an agent based simulation model that was calibrated to survey data for the city of Portland, USA, and Roth et al. (2011), who used smart card data for the London Underground. Roth et al. (2011) further identified that this heterogeneity represented a polycentric organisation of demand with multiple destination hubs, rather than a monocentric organisation in which most trips are to a dominant city centre.

Although none of these studies focus exclusively on travel demand in urban road networks, there is little reason to suppose that the demand distribution should be dramatically different to the heterogeneous distribution that have been described by these studies. However, further empirical studies are required to test this hypothesis. A more detailed understanding of the structure of demand in urban road networks should emerge as new datasets and methods become available.

### 2.2 Numerical Studies of the effects of Network Structure on Performance

Numerical studies of the effects of network structure on performance have, thus far, predominantly focused on the effects of supply structure, rather than demand structure. The typical approach taken by such studies has used canonical models from the Network Science literature to generate large ensembles of synthetic networks, each of which have different structural characteristics. Frequently used models include the random graph model of Erdös and Rényi (1959), the scale-free network model of Barabasi and Albert (1999) and the small-world network model of Watts and Strogatz (1998); for which illustrative examples are shown in Figure 1. The performance characteristics of these different ensembles of networks are then compared, under an assignment of travel demand using a road traffic model, in order to determine which ensemble has the best performance, which is taken as an average across all networks from within each ensemble.

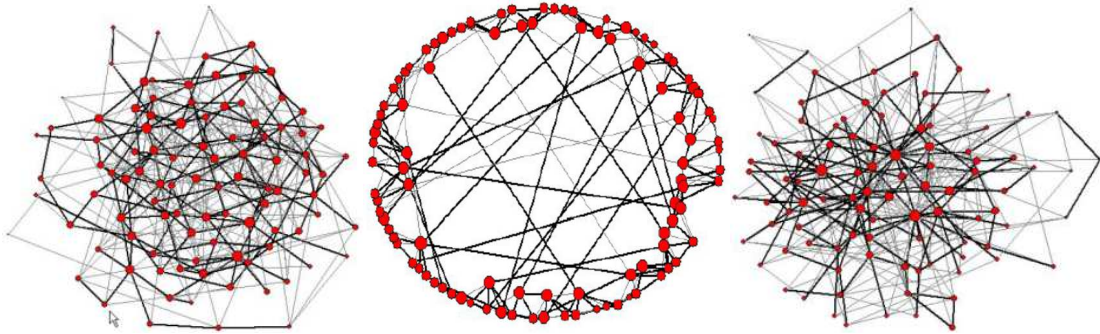

**Figure 1 - Examples of (from left to right) Random, Small-World and Scale-Free Networks (Youn et al., 2008)**

| Study | Topologies | Network Size ($n$ nodes, $m$ links, $\langle k \rangle$: Average Node Degree) | Number of Network Realisations | Link Travel Time Functions $t_i$ ($t_{0i}$: free-flow travel time, $x_i$: link flow, $cap_i$: link capacity) | Demand Structure | Main Performance Indicator |
|---|---|---|---|---|---|---|
| Wu et al. (2006) | Random; Scale-Free; Small World | $n = 400$; $m = 1400$; $\langle k \rangle = 7$ | 25 | $t_i = t_{0i}\left[1 + 0.15\left(\frac{x_i}{cap_i}\right)^4\right]$<br>- $t_{0i} \in (0, 0.1]$ randomly selected for each link<br>- $cap_i \in [20, 60]$ randomly selected for each link | Random | Proportion of links over Capacity |
| Zhao and Gao (2007) | Regular Ring; Random; Scale-Free; Small World | $n = 500$; $m = 1000$; $\langle k \rangle = 4$ | 50 | $t_i = t_{0i}\left[1 + 0.15\left(\frac{x_i}{cap_i}\right)^4\right]$<br>- $t_{0i} \in (0, 1]$ randomly selected for each link<br>- $cap_i = 10000$ for each link | Uniform | Total Travel Time |
| Youn et al. (2008) | 1D Regular Lattice; Random; Scale-Free; Small World | $n = 100$; $m = 300$; $\langle k \rangle = 6$ | 50 | $t_i = a_i + b_i x_i$<br>- $a_i \in \{1, 2, 3\}$ randomly allocated to each link<br>- $b_i \in \{1, 2, \ldots, 100\}$ randomly allocated to each link | Single OD pair | Price of Anarchy |
| Wu et al. (2008a) | Random; Scale-Free | $n = 100$; $m = 1350$; $\langle k \rangle = 2.7$ | 100 | $t_i = t_{0i}\left[1 + 0.15\left(\frac{x_i}{cap_i}\right)^4\right]$<br>- $t_{0i} \in (0, 0.1]$ randomly selected for each link<br>- $cap_i = C \ \forall i$ but the value of $C$ is not defined | Not defined | Proportion of links over capacity |
| Wu et al. (2008b) | Regular Lattice; Random; Scale-Free; Small World | $n = 100, \ldots, 1000$; $m = 100, \ldots, 1000$; $\langle k \rangle = 2$ | 50 | $t_i = t_{0i}\left[1 + 0.15\left(\frac{x_i}{cap_i}\right)^4\right]$<br>- $t_{0i} \in (0, 1]$ randomly selected for each link<br>- $cap_i$ is not defined | Random | Price of Anarchy |
| Sun et al. (2012) | Scale-Free with variable community structure | $n = 100, 160, 220$; $m = 400, 640, 880$; 4 communities | 20 | $t_i = t_{0i}\left[1 + 0.15\left(\frac{x_i}{cap_i}\right)^4\right]$<br>- $t_{0i}$ randomly selected for each link<br>- $cap_i = 60 \ \forall i$ | Random | Proportion of links over capacity |
| Zhu et al. (2014) | Scale-Free; Small World | $n = 1000$; $m = 3000$; $\langle k \rangle = 6$ | Not defined | $t_i = t_{0i}\left[1 + 0.15\left(\frac{x_i}{Ce_i}\right)^4\right]$<br>- $t_{0i} = 1$ for every link $i$ between nodes $i1$ and $i2$<br>- $Ce_i = \min(Cn_{i1}/k_{i1}, Cn_{i2}/k_{i2})$, for which i) $Cn_j$ is fixed and ii) $Cn_j = f(k_j)$ | Uniform; Gravity Model | Volume to Capacity ratio (V/C) |

**Table 3 - Summary of Network Science Studies of the effects of Network Structure on Network Performance**

Table 3 describes seven numerical studies that used this synthetic networks approach, alongside traffic equilibrium modelling techniques, to study the variation of several different performance indicators as total demand was increased in a variety of different network ensembles. The columns in this table illustrate that these studies used a broad range of supply and demand structures. On the supply-side, significant variation is evident with respect to the numbers of nodes and links used, and also the parameter settings for link cost coefficients, which, in most cases, were chosen either randomly, from within a given range, or were fixed at one value, which was then applied to each link in each network. Significant variation is also evident on the demand-side. For example, Wu et al. (2006), Wu et al. (2008b) and Sun et al. (2012) used a *random* structure of demand wherein, as travel demand increased, each increment of total demand was wholly allocated to a randomly selected origin-destination node pair. Whereas, in contrast, Zhao and Gao (2007) and Zhu et al. (2014) used a uniform structure of demand in which each increment of total demand was spread evenly across all origin-destination node pairs.

The studies described in Table 3 all found that network performance does indeed vary with respect to supply and demand structure. However, their findings are inconsistent. For example, whilst Youn et al. (2008) found that scale-free networks performed the best, followed by random and lattice networks, and that small-world networks performed the worst, Wu et al. (2008b) found a different ordering in which scale-free networks performed the best, followed by small-world and random networks, and that lattice networks performed the worst. These differences presumably arise from differences in the selected configurations of supply and demand structure, but it is difficult to gain much insight from considering these two studies together because so many aspects of network structure are different between them.

From a broader perspective, the studies presented in Table 3 do not provide a clear connection with the empirical studies described in section 2.1. Indeed, all of the studies used structures of supply that are not plausible for urban road networks because they do replicate the structural features observed in real networks. With respect to topological structure, as is illustrated by Figure 1, the random, small-world and scale-free graph models typically produce non-planar networks. Moreover, the values chosen for link free-flow travel costs $t_{0i}$ in each study indicate that the networks would not allow a geographical embedding.

From a methodological perspective, it is not clear that a comparison of the average performance of each ensemble of networks is appropriate because none of these studies provide justification for whether this is a suitable summary statistic. Such justification would require discussion of the distribution of performance across network realisations within each ensemble, but this analysis is only undertaken by Youn et al. (2008). It is also highlighted that three of the seven studies in Table 3 do not provide complete descriptions of all of the parameter settings used; specifically the studies of Wu et al. (2008a), Wu et al. (2008b) and Zhu et al. (2014). These studies cannot, therefore, be reproduced and independently verified by other researchers.

These issues highlight the need for an experimental framework such at that proposed in this paper.

# 3 The Proposed Experimental Framework

## 3.1 The Main Challenge: Selecting Networks from the Search Space

The review in section 2.2 illustrates that investigations of the effects of network structure on performance have to contend with a huge, multi-dimensional search space of networks, which spans all possible configurations of supply and demand structure. This is particularly well illustrated by the columns in Table 3. On the supply-side, there is a huge, multi-dimensional space of possible infrastructure configurations; with respect to the numbers of nodes and links, their connection pattern and the functional form and associated parameters of how link costs are represented. Similarly, there is also a broad array of possible demand patterns; with respect to both the total amount and the distribution of demand between nodes in the supply network.

Given this high dimensionality, the challenge faced by any numerical investigation is how to select network configurations for comparison from within this search space so as to provide insights into how network structure affects performance in urban road networks. The selection of network configurations is important because it has a direct impact on the strength, generality and transferability of research findings. There are two approaches to network selection that been used in existing literature: the synthetic networks approach of the Network Science papers, and the real-world data approach of Parthasarathi and Levinson (2010) and Levinson (2012), mentioned in section 1. We address each of these approaches in turn.

### 3.1.1 Approach 1: The Synthetic Networks Approach

As described in the review in section 2.2, the Network Science approach uses a small number of canonical models from the Network Science literature to generate large ensembles of synthetic networks whose performance characteristics are then compared. A visual illustration of how this approach selects networks from the search space is shown in Figure 2. In this figure, the square represents the search space[3] and the dashed line marks the boundary between planar and non-planar supply networks. Three ensembles of networks are illustrated in the non-planar region, which could represent, for example, scale-free, small-world and random networks.

---

[3] Figure 2, Figure 3 and Figure 4 are used only to illustrate the differences between the different approaches to network selection. They should not be interpreted as accurate representations of the search space.

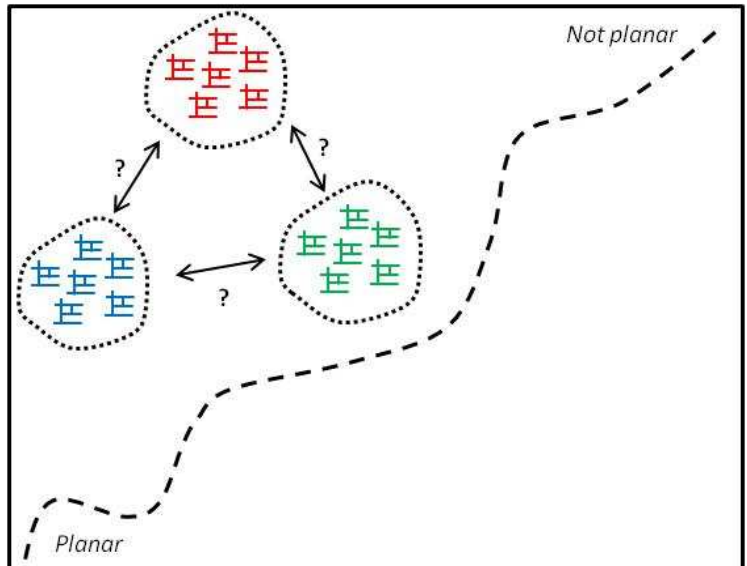

**Figure 2 – Illustration of network selection from the search space for the Network Science approach**

One of the main issues with this approach, which has already been identified, is that it has typically been applied to generate supply networks that are non-planar and are, therefore, not plausible as representations of real urban road networks. This method of selecting networks from the search space has an additional flaw in that it provides only point-to-point comparisons between ensembles of particular network types, whose similarity/dissimilarity in structure is unclear. This is illustrated by the arrows and question marks in Figure 2. For example, the separate categories of scale-free, small-world and random networks used by Wu et al. (2006) are compared with each other, but these comparisons give little insight into networks that do not neatly fit within these categories. As a result, this approach does not explain which aspects of supply or demand structure cause the differences in performance that have been observed, and is capable only of providing results of the form: networks of type 'A' perform better, on average, than networks of type 'B'.

### 3.1.2 Approach 2: The Real-World Data Approach

An alternative network selection technique is demonstrated by Parthasarathi and Levinson (2010) and Levinson (2012). In this second approach, real data for a large sample of real urban road networks are used, alongside regression techniques, to search for correlations between measures of supply and demand structure and indicators of network performance. A visual illustration of this second approach is shown in Figure 3. In the context of selecting networks from the search space, this approach focuses on networks from the planar region and uses a larger sample of different network types than is typically used in the first approach, thereby providing greater coverage of the search space. In selecting networks in this way, this approach provides a large number of individual readings for a range of structural measures, which can each be paired with the value of a performance indicator, thereby enabling correlations between these variables to be studied. For example, Levinson (2012) studied the correlation between fifty values of the typical journey to work travel time in fifty US cities and the population size of those fifty cities.

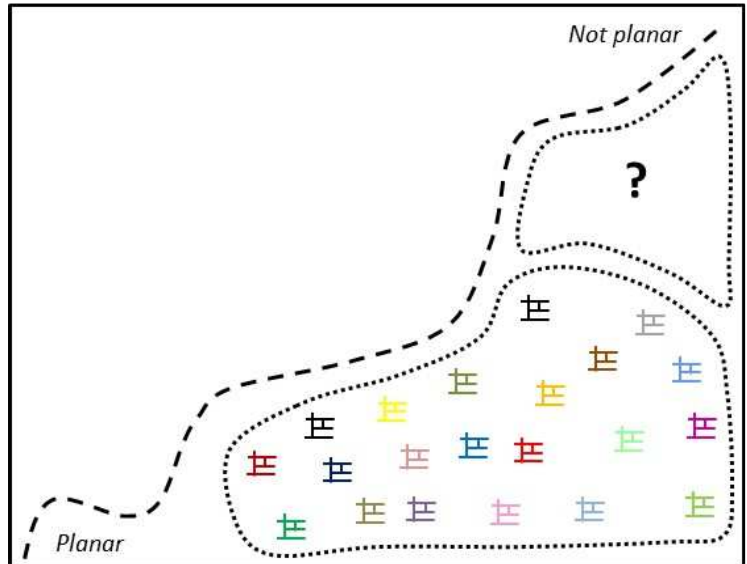

**Figure 3 – Illustration of network selection from the search space for the Levinson (2012) approach**

This approach also has drawbacks. Firstly, it is data intensive as it requires data on network supply, travel demand and network performance for each selected city, which may not be readily available or of sufficient quality. For network performance data, Parthasarathi and Levinson (2010) and Levinson (2012) used publically available data from the Texas Transportation Institutes Urban Mobility Report (Schrank et al., 2012). However, such sources are limited to the indicators of interest to the organisation that collected the data and are also subject to noise as a result of the way in which the data has been collated and processed. This approach is also restricted to those networks for which the required data are available, which, in the context of the huge number of possible supply and demand configurations, are unlikely to span the range of interest for possible structures of urban road networks. This is illustrated by the question mark region in Figure 3. The networks in each sample are also likely to be different from each other in aspects of supply and demand structure, and such variation cannot be controlled. These drawbacks restrict the capability of such regression analyses to identify which aspects of structure impact upon observed variations in performance. This is perhaps one reason for the small R-squared values reported by both Parthasarathi and Levinson (2010) and Levinson (2012) for the explanatory power of their statistical models.

### 3.1.3 The Proposed Approach: Fusion of Synthetic Networks and Real-World Data

The approach to network selection proposed in this paper combines the benefits of flexibility and control in network generation, which is offered by the synthetic networks approach of Network Science, with the idea of studying an array of networks that span a range real urban road network structures, which is the basis of the regression analyses in the approach of Parthasarathi and Levinson (2010) and Levinson (2012). The proposed approach is to generate a spectrum of ensembles of synthetic networks, that provide a cross-section of the search space, in which only one aspect of network structure is varied, and to then use a road traffic model to explore how performance indicators vary within each ensemble and with respect to the selected aspect of network structure. This approach is illustrated in Figure 4. Within this approach, the aspects of structure that do not vary are fixed at values or in a configuration that is plausible for real urban road networks, whilst the aspect of structure that varies, does so to encompass a range of values for that structural feature that have been observed in real urban road networks. A simple example of this approach would be a spectrum of network ensembles in which total demand is increased by a global demand multiplier, whilst network supply and distribution of travel demand are fixed in plausible configurations.

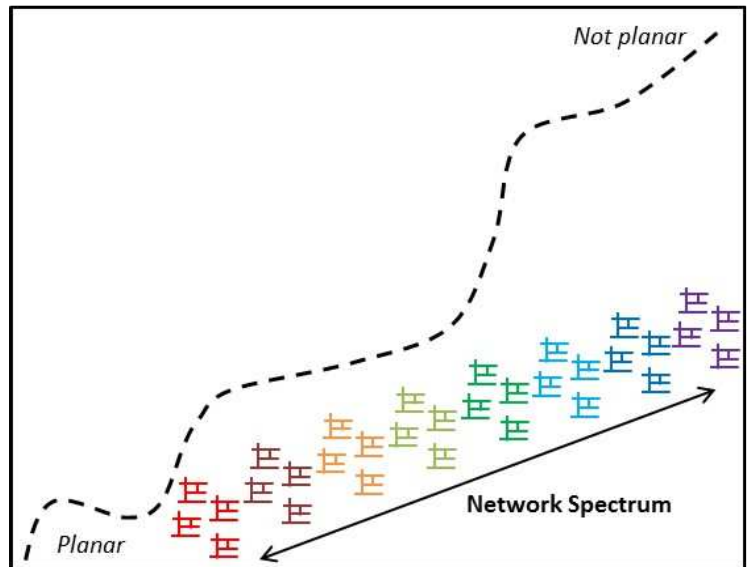

**Figure 4 – Illustration of network selection from the search space for the approach proposed in this paper**

In comparison with the previous approaches, illustrated in Figure 2 and Figure 3, this approach avoids comparisons between distinct categories of networks, which was a key flaw of the first approach, because one aspect of network structure varies 'continuously' across the spectrum of networks. It also avoids the problem of there being too many aspects of network structure that change at the same time, which, in the second approach, made it difficult to establish relationships between measures of structure and performance indicators. Like the second approach, this approach is still reliant on data for supply and demand structure in order to calibrate the generation of synthetic networks. However, unlike the second approach, this approach does not require additional data on network performance because this is produced by the road traffic model.

### 3.2   Statement of the Proposed Framework

The proposed experimental framework comprises five steps:

Step 1.  Identify a measurable aspect of network supply or demand structure and a network performance indicator to be studied.
Step 2.  Design a model of road network generation that produces a spectrum of network ensembles, which span a range of values for the aspect of network structure to be studied and which are plausible as representations of real urban road networks.
Step 3.  Calculate values of the performance indicator for the generated networks using a road traffic model.
Step 4.  Graph the calculated values of the performance indicator against a measure of the selected aspect of network supply or demand structure that has been varied.
Step 5.  Document the preceding steps to include a complete description of the parameter settings used for supply and demand, making the results reproducible.

Empirical studies of the structure of urban road networks are useful for the first and second steps in this framework because they motivate interesting aspects of structure for investigations, and also help to define structural characteristics that generative models of urban road networks should aim to replicate. With reference to the third step, the framework can accommodate any road traffic model of any level of complexity; the only conditions on its specification are that the model should be plausible for road traffic and that it should be adequately described (see step five). The graph produced in step four provides insight into how the selected aspect of network structure affects the selected performance indicator. This graph will also reveal the dispersion of network performance values when the mechanism for network generation has a stochastic component. The final step of the framework ensures that experiments can be reproduced by other researchers, avoiding duplication of tests and providing an intelligble foundation for future work. This documentation should include a description of the road traffic model and, where appropriate, provide data to support the stability and validity of model outputs. Inclusion of such data provides confidence that any presented differences in performance are real. This is especially important when comparisons are to be made between a large number of different networks.

## 4  An Example Application of the Proposed Experimental Framework

In accordance with the first step of the framework and inspired by the empirical studies that were described in section 2.1, the four numerical experiments presented in this section investigate how two performance indicators vary with respect to the density of travel demand and the size, density and connectivity of network supply structure. The two performance indicators are the average link Volume-to-Capacity (V/C) ratio, which is used as a measure of congestion, and the Price of Anarchy, which measures the inefficiency of the selfish behaviour of road users in comparison with socially optimal behaviour (Koutsoupias and Papadimitriou, 1999, Papadimitriou, 2001).

The sections that follow document how the experimental framework has been applied in these four experiments (i.e. the fifth step). Section 4.1 describes the model of road network generation and section 4.2 describes how this model has been used to create spectrums of synthetic network ensembles for each numerical experiment. These two sections thereby address the second and third steps of the framework. The results of the four experiments are then presented in section 4.3 and discussed in section 4.4, thereby addressing the fourth step of the framework.

### 4.1   Model of Road Network Generation

The model described in this section generates a single realisation of a synthetic road network. There are four stages in this generative process: the creation of a topological and geometric structure, described in section 4.1.1; the allocation of travel time functions to network links, described in section 4.1.2; the generation of a travel demand structure, described in section 4.1.3; and the specification of a road traffic model, described in section 4.1.4.

#### 4.1.1   Network Model

There are broadly two approaches that have been proposed as generative models for the topological and geometric structure of road networks. The first approach, used by Barthelemy and Flammini (2009) and Courtat et al. (2011), starts with an initial seed network in a square domain and incrementally adds new nodes to this domain over time, with each node stimulating the growth of new network links to connect to the existing network. The second approach, used by Vitins and Axhausen (2010), assumes a predetermined number and distribution of nodes in the domain, which is used to define a set of candidate links. Network links are then selected from this candidate link set via an optimisation procedure that minimises some objective function, such as total travel cost, subject to a network construction budget.

Both Barthelemy and Flammini (2009) and Courtat et al. (2011) demonstrated that their models were able to reproduce many of the structural features observed in real urban road networks, such as those described in section 2.1.1. Both models also have a range of input parameters and settings that could be used or adjusted to produce a broad range of networks with different structural features. As such, both of these approaches could be used within the proposed framework. However, these models are not freely available to use, and so a simpler model of network generation is proposed instead, which is more similar in approach to the method of Vitins and Axhausen (2010) but which omits the optimisation procedure.

This network model has three steps:

- **Step 1: Scatter $n$ nodes randomly in a square domain, which is $A$km² in size** - As uniformly randomly distributed nodes tend to occur in clusters, which would result in links of extremely short length, a rule is imposed that all nodes must be at least $d_{min}$km apart; i.e. a minimum link length.

- **Step 2: Construct the Minimum Weight Spanning Tree (MST) and Delaunay Triangulation on the node set generated in Step 1** - The Delaunay Triangulation for a set of nodes $V$ is a triangulation of the node set that maximises the minimum angle of all triangles and contains the maximum possible number of links without violating planarity. Euler's formula shows that the maximum number of links in this graph is $3n - 6$ (Barthelemy, 2011). The MST is the graph of minimum total length that provides a path between every pair of nodes for a given node set. It is a subgraph of the Delaunay Triangulation and contains $n - 1$ links. The Delaunay Triangulation and MST define the candidate link set for the new network.

- **Step 3: Select $m \geq n - 1$ links from the Delaunay Triangulation to create a new network $G$ defined on this node set, of which the first $n - 1$ links are from the MST** - The inclusion of the MST in $G$ guarantees that there is at least one path between every pair of nodes, i.e. that the network is fully connected. The remaining $m - (n - 1)$ links are randomly selected from the remaining links in the Delaunay Triangulation.

The number of nodes $n$, the domain size $A$, the minimum link length $d_{min}$ and the number of links $m$ are input parameters to the model, which, when varied, generate networks with a wide variety of structures; see Figure 5 for examples. As the scattering of nodes in step 1 and the selection of links in step 3 are stochastic processes, this model produces a different network each time it is run. Several individual model runs can therefore be used to create ensembles of networks, which have similar structural features.

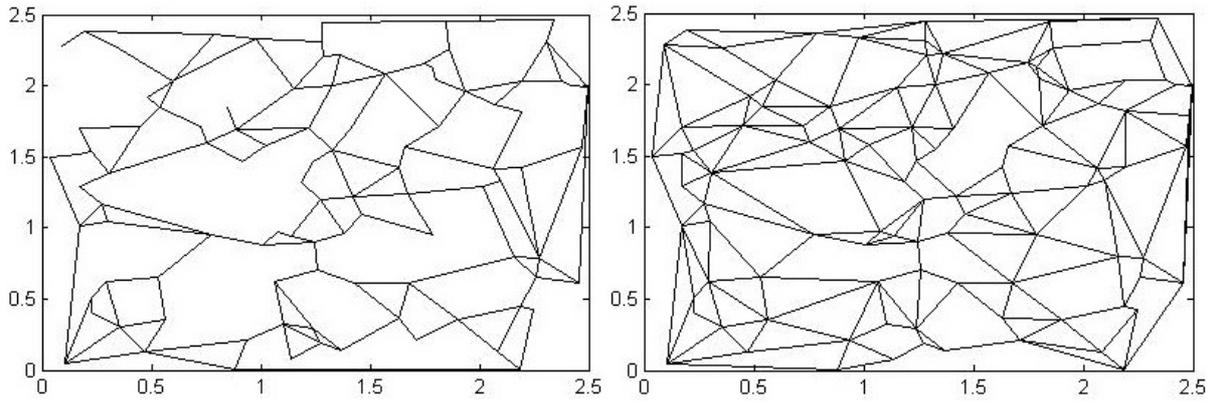

Figure 5 – Two example network realisations with input parameters $n = 100$, $A = 6.25$ and $d_{min} = 0.05$, with $m = 159$ (left) and $m = 229$ (right)

The networks generated by this model do not replicate all of the structural properties of supply in urban road networks identified in section 2.1. For example, Figure 5 illustrates that typical networks produced by this model have a wide range of angles between links at nodes. However, the networks generated by this model are planar and are therefore more plausible for urban road networks than the networks used in the studies described in section 2.2.

### 4.1.2 Link Travel Time Functions

Link travel time functions describe the time required to traverse a given link as a function of the volume of flow and its operational characteristics. A commonly used link travel time function is the BPR function (Bureau of Public Roads, 1964), which has the following form:

$$t_i = t_{i0}\left[1 + 0.15\left(\frac{x_i}{cap_i}\right)^4\right]$$

where, for a link $i$, $t_{i0}$ represents the travel time in free-flow conditions, $x_i$ represents link flow and $cap_i$ represents link capacity. As this function is commonly used in the transportation literature and also the studies referenced in Table 3, it is also adopted in this model for each network link.

In general, the values of $t_{i0}$ and $cap_i$ depend on factors that include link length, the speed limit, the level of street frontage activity and the number of lanes. Guidance published for modelling practitioners by the Department for Transport in the UK recommends an assumption that links in central urban areas - where there is a speed limit of 30mph (48kph) - should have a maximum modelled capacity of $Q_i = 800$ vehicles per hour, per 3.65m lane (WebTAG, 2014). Adopting this form, for each link $i$, we set $cap_i = Q_i \times k_i$, where $k_i$ represents the number of lanes, and $t_{i0} = d_i/48$, where $d_i$ represents link length in kilometres. The units of $t_{i0}$ are therefore hours. The number of lanes on each link $k_i$ and the lane capacity value $Q_i$ are input parameters to the model. The length $d_i$ for each link is generated by the network model described in section 4.1.1.

### 4.1.3 Travel Demand

In lieu of adequate empirical data on the structure of demand in urban road networks (see section 2.1.2), this paper makes a simple assumption that total travel demand is uniformly distributed across all node pairs in each network. Each node pair $r$ is therefore assumed to have the same amount of demand $q_r$ travelling between them. This approach was used by studies described in Table 3.

The total amount of demand in each network is set as a function of the domain size $A$ and is controlled so that a constant density of demand per km², $\rho_{dem}$, can be maintained across network domains of different sizes. The demand density $\rho_{dem}$ is set as an input parameter to the model. The generation of demand in this way is one way of ensuring that performance comparisons are relatively fair between networks that serve domains of different sizes. This approach includes implicit assumptions that demand density does not vary within the domain or as the size of the domain changes.

It follows from these assumptions that the amount of demand per OD pair $r$ is:

$$q_r = \frac{A\rho_{dem}}{n(n-1)}$$

### 4.1.4 Road Traffic Model

The final stage of the model is the assignment of travel demand from section 4.1.3 to each supply network created in sections 4.1.1 and 4.1.2. In this paper, travel demand is assigned in accordance with two different principles of route choice; the User-Equilibrium (UE) and System Optimum (SO) routing principles (Wardrop, 1952). Under the UE principle, individual travellers choose routes such that they each, selfishly, minimise their individual travel costs. Under the SO principle, in contrast, individual travellers choose routes such that the Total Travel Cost across all travellers is minimised. The Price of Anarchy measures the difference between these two principles and is calculated as the ratio of the Total Travel Cost under UE to the Total Travel Cost under SO (Roughgarden, 2005).

Subject to conditions on link travel time functions, which those described in section 4.1.2 satisfy, and other consistency conditions, it can be shown that there exist unique link flows $x_i^{UE}$ and $x_i^{SO}$ satisfying the UE and SO principles respectively (Sheffi, 1985). Due to the complexity of traffic assignment problems, traffic flows under UE and SO are typically derived using numerical methods. All of the numerical results presented in this paper were generated by the Origin Based Assignment (OBA) algorithm (Bar-Gera, 2002), using the OBA executable from Bar-Gera (2001). Each traffic assignment is solved to an Average Excess Cost of at most $10^{-4}$, which accords with the guidance on convergence provided by Boyce et al. (2004).

### 4.2 Description of Numerical Experiments and Road Traffic Model

The model of road network generation, described in section 4.1, has seven input parameters in total. Table 4 describes the values used for four of these parameters ($A$, $n$, $m$ and $\rho_{dem}$) to create spectrums of network ensembles across the four structural dimensions of interest. This table also displays corresponding values for node density $\rho_n$ and meshedness $M$, which measure the density and connectivity of network supply structure.

In the first experiment, given the absence of empirical data on travel demand, demand density was varied across a broad range of values between $\rho_{dem} = 1250$ and $\rho_{dem} = 7950$. Whilst demand density was varied, the parameters that control the density and connectivity of network supply were fixed at average observed values for real urban road networks, taken from Table 1 and Table 2 respectively. The domain size $A$ and number of nodes $n$ were also fixed in this experiment, although at much smaller values than the average observed values of $A = 296$km² and $n = 4713$ shown in Table 1. This is because a road traffic model run in a network of such size represents a significant computational burden; for example, a single run of the road traffic model described in section 4.1.4, for a network with input parameters $A = 296$, $n = 4713$, $m = 7538$, $\rho_{dem} = 2355$, $d_{min} = 0.05$,

$k_i = 1$ and $Q_i = 800\ \forall i$, took approximately 36 hours to find a solution of sufficient precision. Repetition of this run time over a large number of network realisations would be impractical.

The second, third and fourth experiments, which explore the effects of network size, network density and network connectivity, were setup in a similar way, with one aspect of network structure varying whilst the remaining aspects remained unchanged. The range of network sizes in the second experiment was capped at $n = 500$ nodes in order to limit the computational burden of the experiments. The ranges of network density and network connectivity used in the third and fourth experiments encompass the full range of observed values for real urban road networks from Table 1 and Table 2 respectively. In each of the last three experiments, demand density was fixed at $\rho_{dem} = 4350$ because, as will be shown, this value produced a reasonably congested (but not overly-congested) network and also the highest values of the Price of Anarchy.

| Experiment Title | Domain Size ($A$km²) | Num. of Nodes ($n$) | Node Density ($\rho_n$) | Num. of Links ($m$) | Meshedness ($M$) | Demand Density ($\rho_{dem}$) |
|---|---|---|---|---|---|---|
| 1. Demand Density | 6.25 | 100 | 16 | 158 | 0.3 | 1250, 1300, …, 7900, 7950 |
| 2. Network Size | 1.25, 1.875, …, 30.625, 31.25 | 20, 30, …, 490, 500 | 16 | 30, 46, …, 782, 798 | 0.3 | 4350 |
| 3. Network Density | 6.25 | 20, 25, …, 295, 300 | 3.2, 4, …, 47.2, 48 | 30, 38, …, 470, 478 | 0.3 | 4350 |
| 4. Network Connectivity | 6.25 | 100 | 16 | 99, 104, …, 284, 289 | 0, 0.03, …, 0.95, 0.97 | 4350 |

Table 4 - Parameter Settings for Numerical Experiments

Values for the three remaining parameters ($d_{min}$, $k_i$ and $Q_i$) were fixed in all four experiments. The minimum link length was set at $d_{min} = 0.05$km, which coincides with the shortest length of city blocks in the urban areas studied by Chan et al. (2011). The number of lanes was fixed at $k_i = 1$ for each network link, which was based on anecdotal evidence from the UK that one lane per link is the most common situation on urban roads. Finally, link capacity was fixed at $Q_i = 800$ for each network link, which is equal to the maximum value recommended by WebTAG (2014). Coupled with the assumption that every link has one lane, the implication of this assumption is that every network has only one road type in its road hierarchy.

The results presented in the next section are based upon one hundred network realisations for each ensemble of parameter settings shown in Table 4, which corresponds to one hundred individual runs of the network model described in section 4.1. The four spectrums described in Table 4 encompass 135, 49, 57 and 39 separate network ensembles. When combined, the four experiments therefore include results from 28,000 network realisations and 56,000 traffic assignments. All of the traffic assignments were undertaken on a remote desktop server, running the Windows Operating System with 64GB of RAM. The traffic assignment runs for each experiment took 107.75 hours, 421 hours, 67.5 hours and 9.5 hours respectively to complete. This is an overall total run time of 605.5 hours (~25 days)[4].

---

[4] As a caveat to this figure, it is noted that the server used was shared with other researchers at the university, and that this may have increased the run times of our experiments.

## 4.3 Results of Numerical Experiments

The results for the two performance indicators are presented, in each experiment, as a sequence of boxplots for each network ensemble. Data points marked by circles represent network realisations in which the indicator value is greater than the upper quartile, or less than the lower quartile, by between 1.5 and 3 times the interquartile range. Data points marked by stars represent network realisations in which the indicator value is greater than the upper quartile, or less than the lower quartile, by at least 3 times the interquartile range.

### 4.3.1 Experiment 1: Demand Density

Figure 7 and Figure 8 present the variation of the average link V/C ratio and the Price of Anarchy as demand density is increased across the range $\rho_{dem} \in [1250, 7950]$. Supply structure is fixed in each network ensemble, with domain size $A = 6.25\text{km}^2$, $n = 100$ nodes, a node density of $\rho_n = 16$ and a meshedness value of $M = 0.3$. Two examples of such networks are shown in Figure 6.

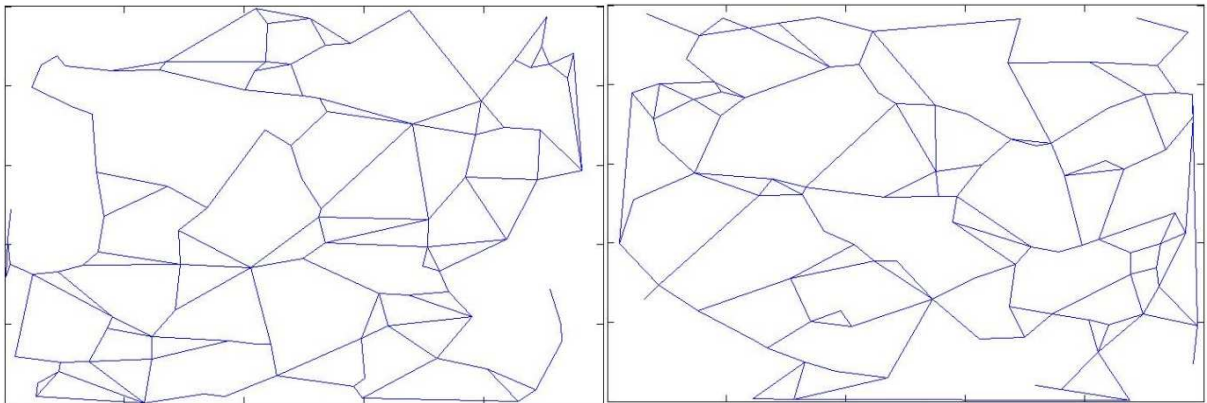

**Figure 6 – Two Example Network Realisations from Experiment 1**

With respect to the level of congestion, Figure 7 shows, perhaps unsurprisingly, that the average link V/C ratio increases monotonically as demand density increases. This increase also appears to be linear. It is hypothesised that this linearity is related to the averaging process within this performance measure. It is also highlighted that the level of dispersion of performance values, across networks within each ensemble, increases as demand density increases, and that the distribution within most ensembles is positively skewed.

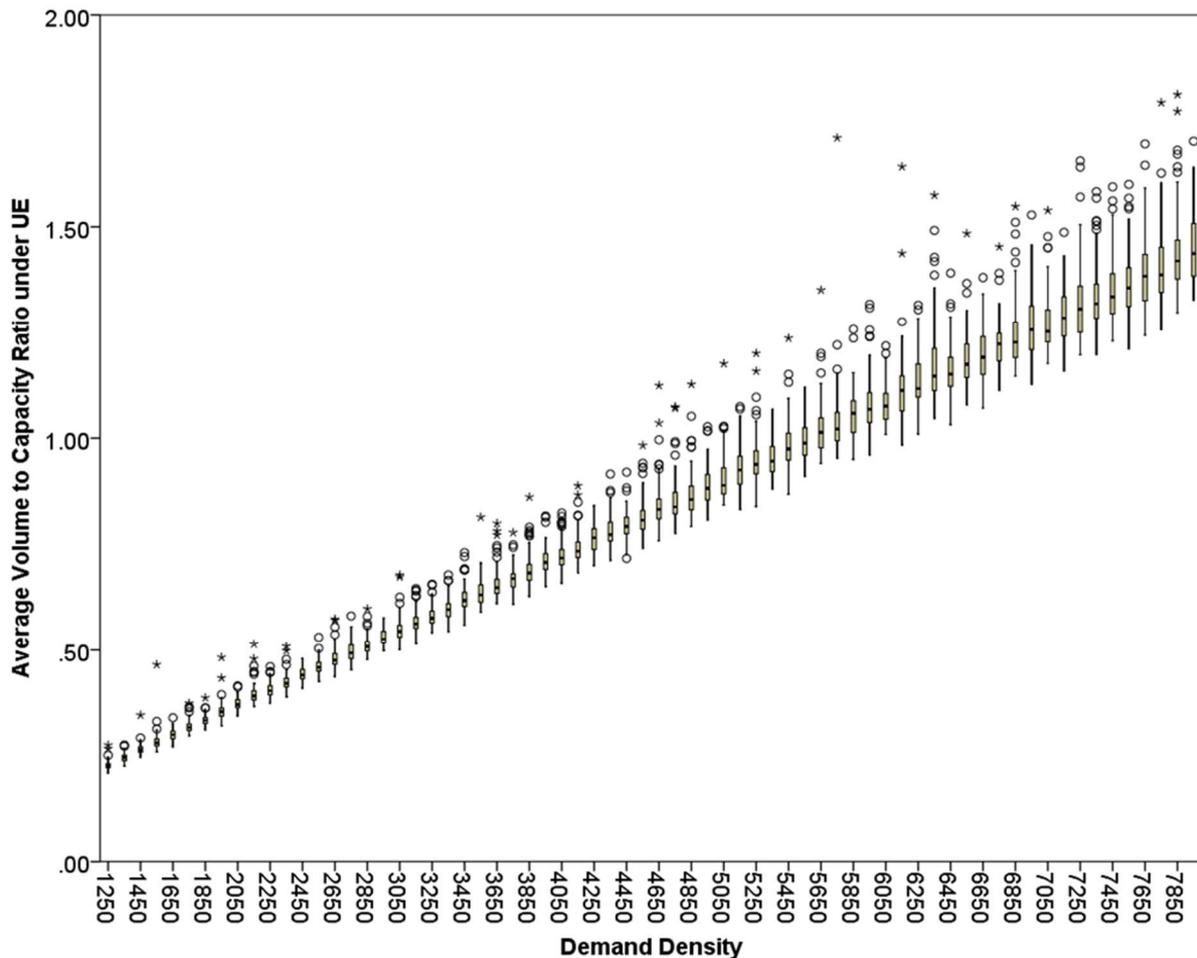

Figure 7 - Average link V/C ratio against Demand Density $\rho_{dem}$ ($A = 6.25, n = 100, \rho_n = 16, m = 158$ and $M = 0.3$)

Figure 8 shows that the Price of Anarchy follows a unimodal pattern as demand density increases; with values initially increasing, before reaching a peak and then falling. The same broad pattern in values of the Price of Anarchy was uncovered by Youn et al. (2008) for increasing demand in random, scale-free, small-world and lattice networks. The dispersion of values of the Price of Anarchy within each ensemble appears to be much wider than is apparent in Figure 7, which suggests that the Price of Anarchy is more sensitive to stochastic variations in network structure. Figure 8 also shows that the level of dispersion within each ensemble initially increases as demand density increases, peaks and then decreases as values of the Price of Anarchy begin to fall.

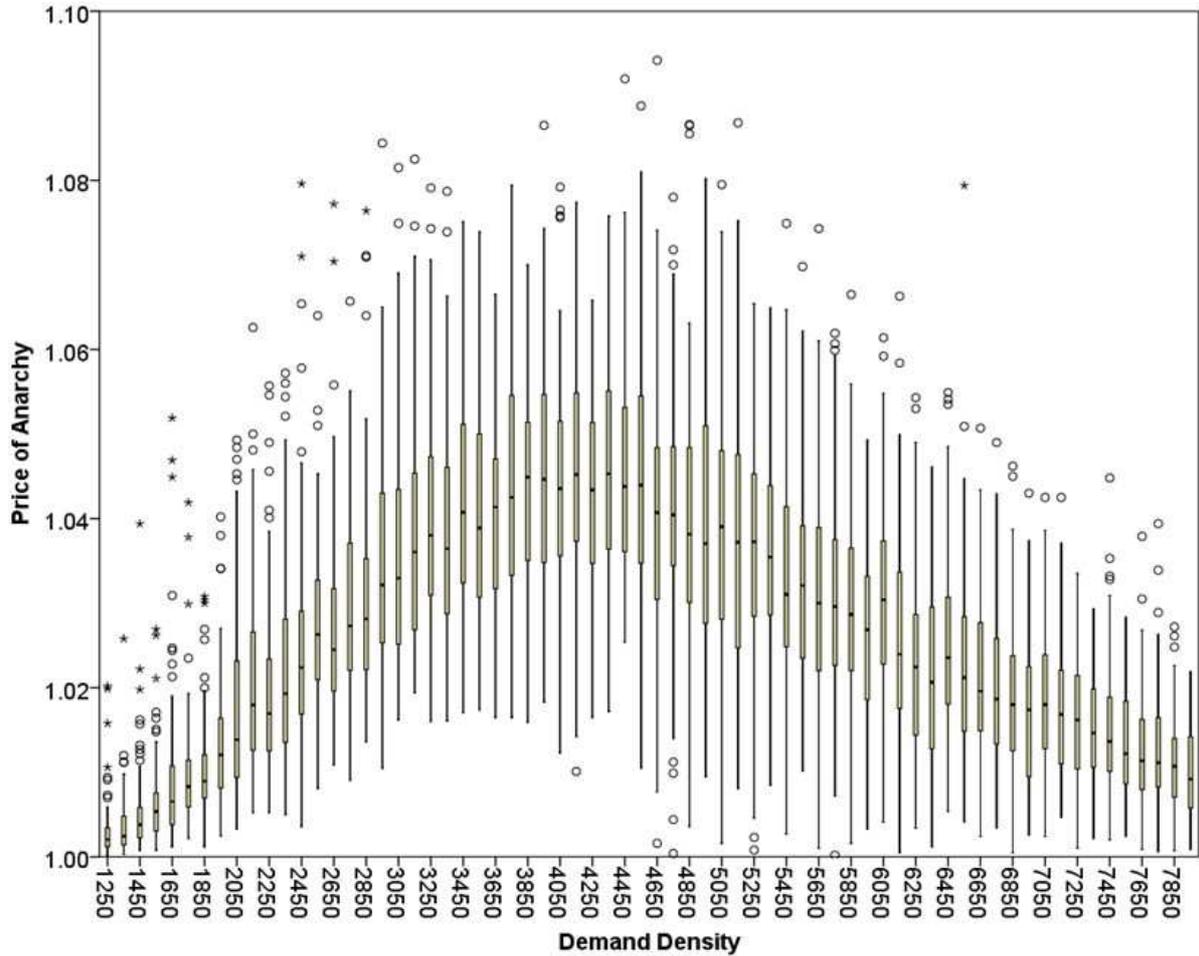

**Figure 8 - Price of Anarchy against Demand Density $\rho_{dem}$ ($A = 6.25, n = 100, \rho_n = 16, m = 158$ and $M = 0.3$)**

### 4.3.2 Experiment 2: Network Size

Figure 10 and Figure 11 present the variation of the average link V/C ratio and the Price of Anarchy as network size is increased, between an ensemble of networks with $n = 20$ nodes in a domain size of $A = 1.25\text{km}^2$ and an ensemble of networks with $n = 500$ nodes in a domain size of $A = 31.25\text{km}^2$. Typical examples of networks at these two extremes and in the middle of the spectrum are shown in Figure 9. In each ensemble, node density is fixed at $\rho_n = 16$, meshedness is fixed at $M = 0.3$ and demand density per km² is fixed at $\rho_{dem} = 4350$. Note that although the *density* of travel demand is fixed, *total* travel demand still increases as the size of the network increases across this spectrum of networks; this is due to the assumptions described in section 4.1.3. As a consequence of these assumptions, the ratio of total travel demand $\sum_r q_r$ to total network supply $\sum_i m \times k_i \times Q_i$ is the same across each network ensemble in the spectrum.

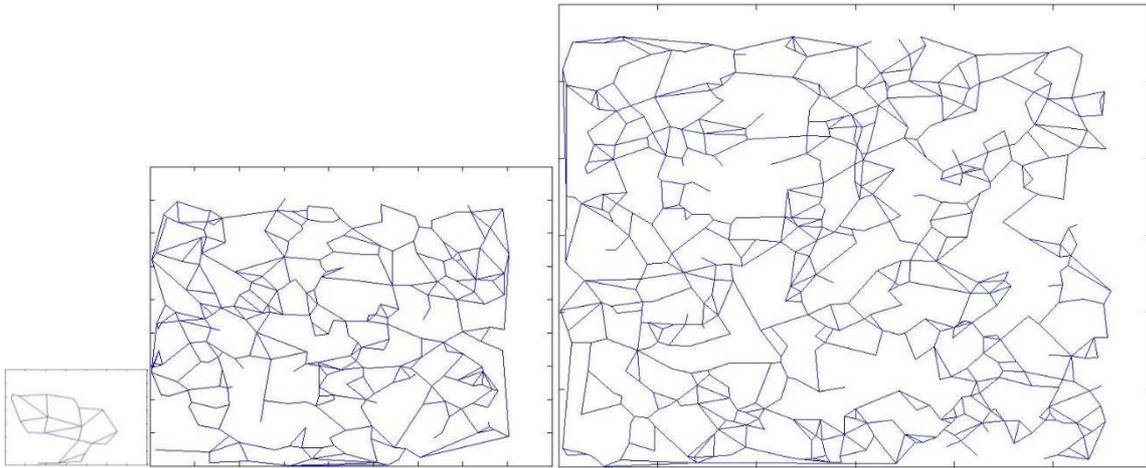

**Figure 9 - Example Network Realisations from Network Ensembles at the lower end (left), in the middle (centre) and at the upper end (right) of the Network Spectrum in Experiment 2**

With respect to congestion, Figure 10 shows that the average link V/C ratio increases monotonically as network size increases. It is also highlighted that the level of dispersion across networks within each ensemble increases as network size increases, and that the distribution within most ensembles is positively skewed.

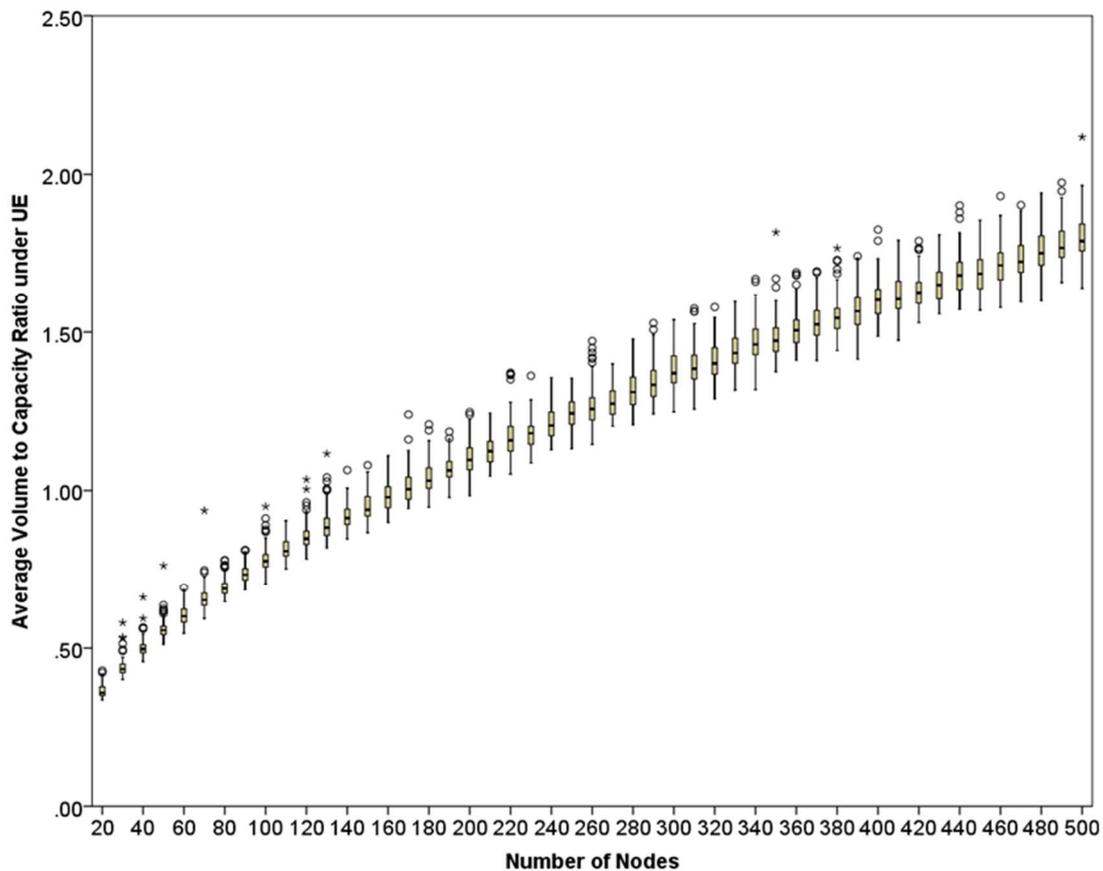

**Figure 10 - Average link V/C ratio against Network Size ($\rho_n = 16$, $m = 158$, $M = 0.3$ and $\rho_{dem} = 2355$)**

Similarly to the results of the first experiment, Figure 11 shows that Price of Anarchy has a unimodal pattern of variation as network size increases. Again the dispersion of values of the Price of Anarchy within each ensemble is wider than is apparent for the average link V/C ratio. The level of dispersion also increases as network size increases and then gradually dissipates, mirroring the broad pattern in the median values of the Price of Anarchy.

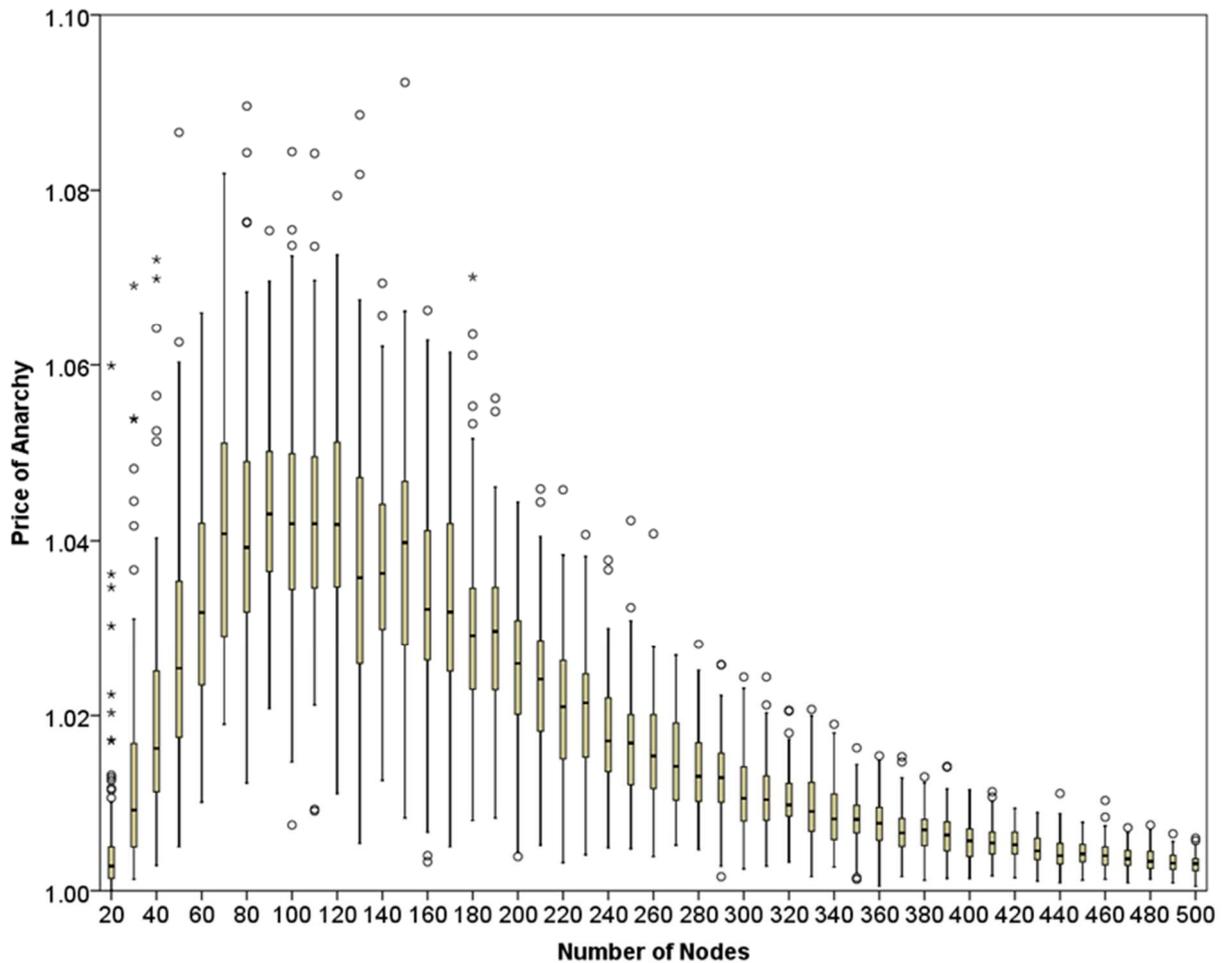

Figure 11 – Price of Anarchy against Network Size ($\rho_n = 16$, $m = 158$, $M = 0.3$ and $\rho_{dem} = 2355$)

### 4.3.3 Experiment 3: Network Density

Figure 13 and Figure 14 present the variation of the average link V/C ratio and the Price of Anarchy as network density is increased, between an ensemble of networks with $\rho_n = 3.2$ nodes per km² and an ensemble of networks with $\rho_n = 48$ nodes per km². Typical network examples at the extremes and in the middle of this spectrum are shown in Figure 12. In each ensemble, the domain size is fixed at $A = 6.25$km², meshedness is fixed at $M = 0.3$ and demand density is fixed at $\rho_{dem} = 4350$. In this experiment, in contrast to experiment 2, the total level of demand remains unchanged because the domain size is fixed. However, the amount of demand per OD pair falls because the number of OD pairs increases as the number of nodes increases across the network spectrum.

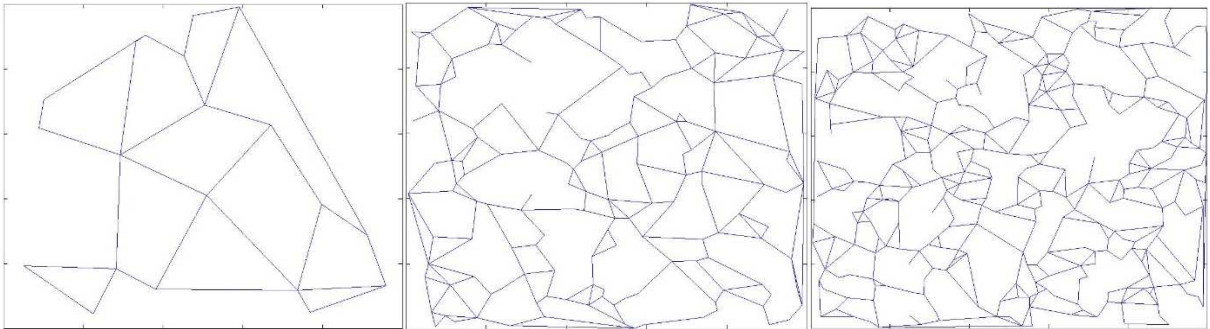

**Figure 12 - Example Network Realisations from Network Ensembles at the lower end (left), in the middle (centre) and at the upper end (right) of the Network Spectrum in Experiment 3**

With respect to the average level of congestion, Figure 13 shows that the average link V/C ratio falls as demand density increases. It is also noted that the levels of dispersion across the networks within each ensemble decrease as network density increases.

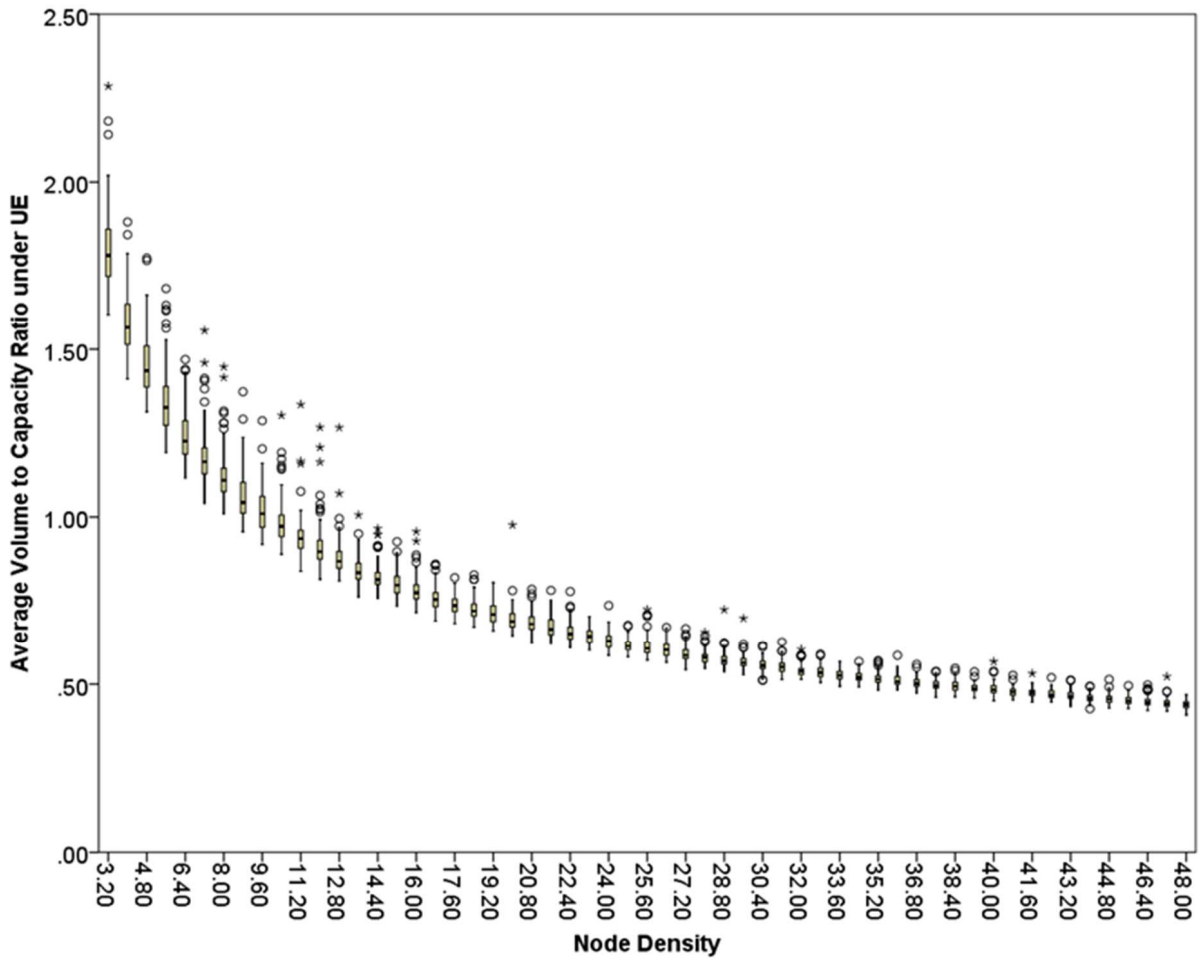

**Figure 13 - Average link V/C ratio against Node Density $\rho_n$ ($A = 6.25$, $M = 0.3$ and $\rho_{dem} = 2355$)**

Similarly to the first two experiments, Figure 14 again shows a unimodal pattern in the Price of Anarchy as demand density increases. However, it is noted that the rate of decay in the Price of Anarchy is much shallower than is shown in either Figure 8 or Figure 11. The levels of dispersion of values of the Price of Anarchy within each ensemble initially increase, as node density increases and then gradually dissipate, again mirroring the overall trend in the Price of Anarchy.

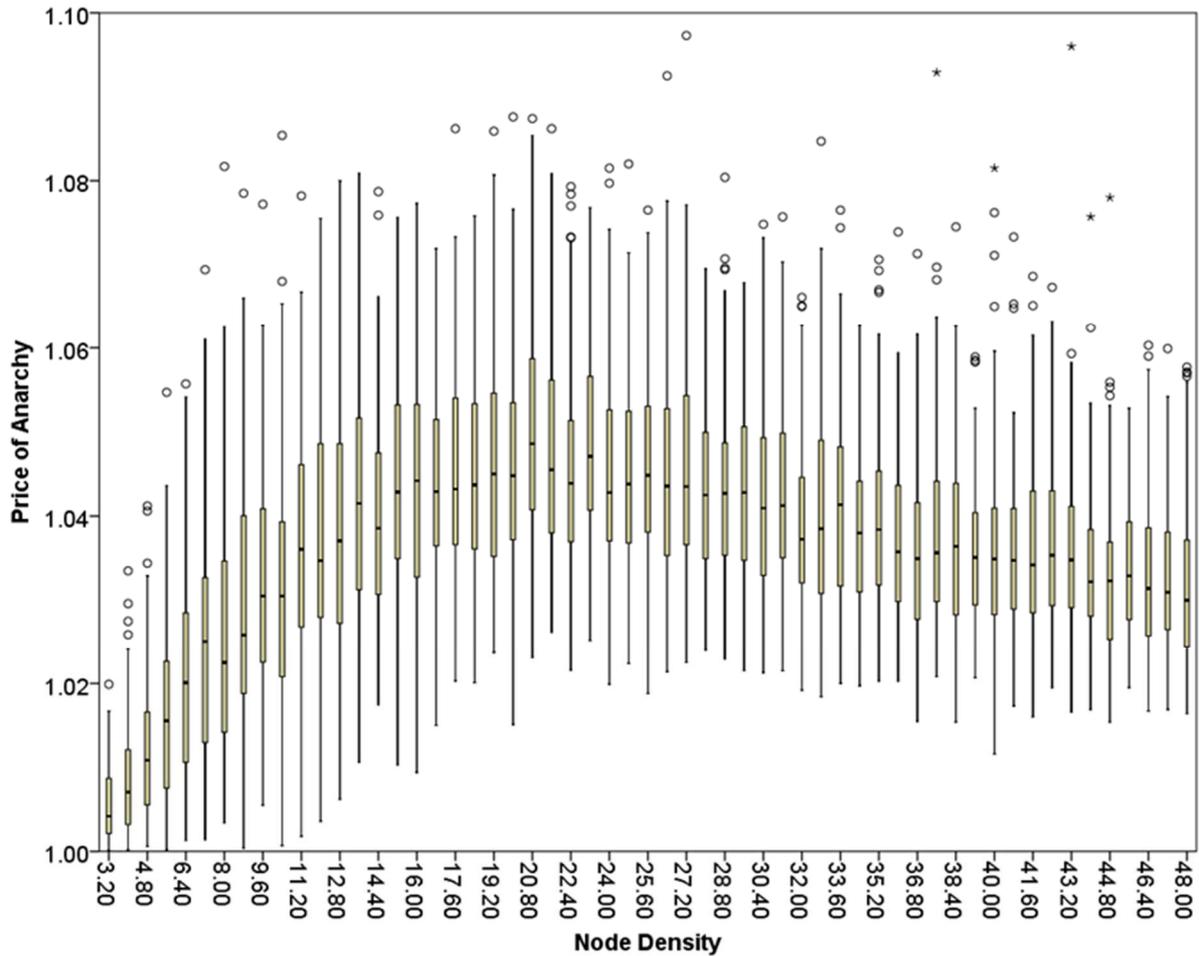

**Figure 14 - Price of Anarchy against Node Density $\rho_n$ ($A = 6.25$, $M = 0.3$ and $\rho_{dem} = 2355$)**

### 4.3.4 Experiment 4: Network Connectivity

Figure 16 and Figure 17 present the variation of the average link V/C ratio and the Price of Anarchy as network connectivity is increased between an ensemble of networks with meshedness $M = 0$ and an ensemble of networks with meshedness $M = 0.97$. Typical network examples of these extremes and a network in the middle of the spectrum are shown in Figure 15. In each ensemble, the domain size is fixed at $A = 6.25\text{km}^2$, the number of nodes is fixed at $n = 100$ and demand density is fixed at $\rho_{dem} = 4350$. In contrast to experiments 2 and 3, the total level of demand and the demand per OD pair both remain unchanged in this experiment.

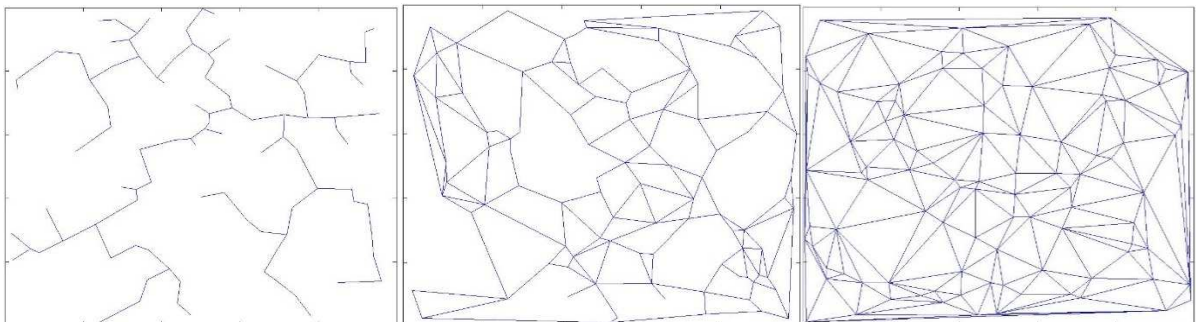

**Figure 15 - Example Network Realisations from Network Ensembles at the lower end (left), in the middle (centre) and at the upper end (right) of the Network Spectrum in Experiment 3**

With respect to the level of congestion, Figure 16 shows that the average link V/C ratio falls as network connectivity increases. Figure 16 also shows that the dispersion of congestion levels across network realisations is significantly different at different levels of connectivity; broadly, dispersion decreases as the level of network connectivity increases.

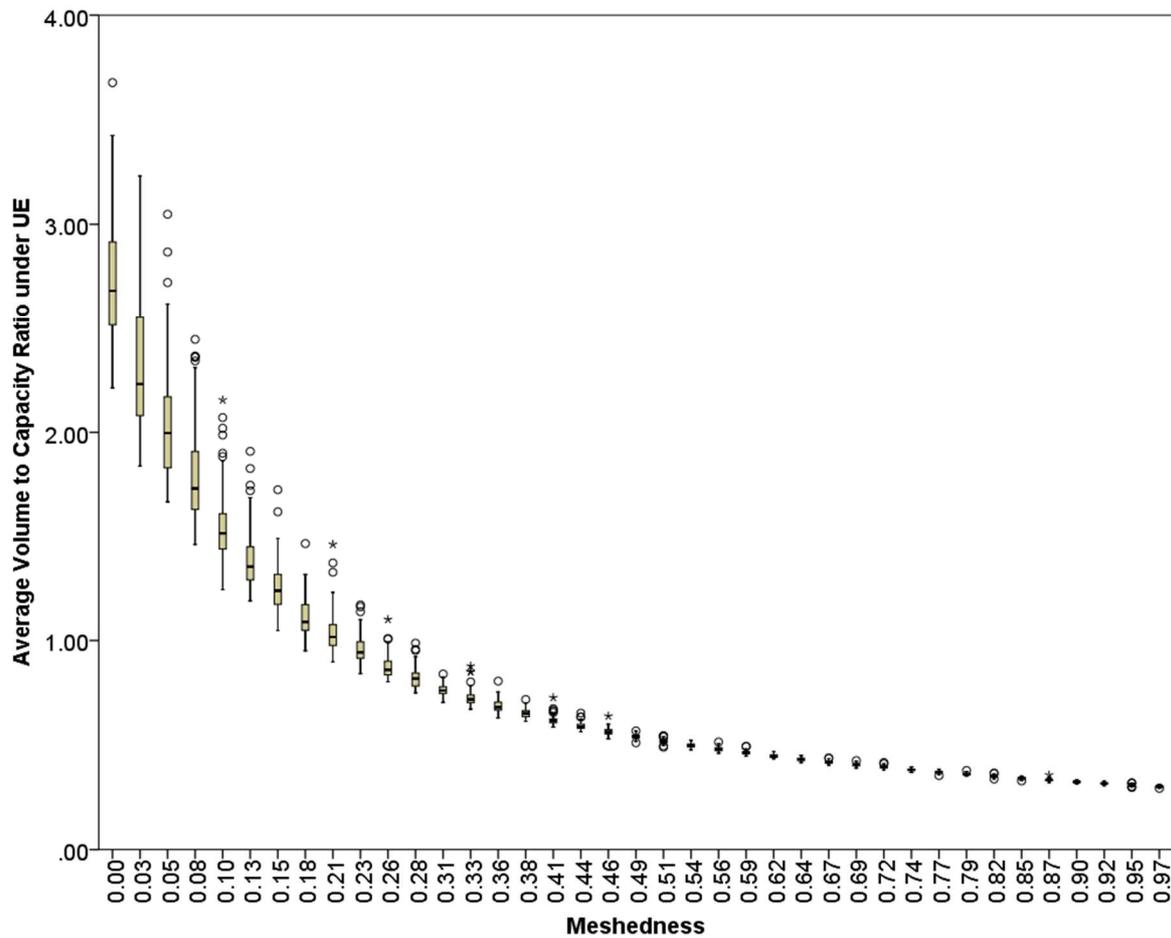

Figure 16 - Average link V/C ratio against Meshedness $M$ ($A = 6.25$, $n = 100$, $\rho_n = 16$ and $\rho_{dem} = 2355$)

Similarly to the three previous experiments, the variation of the Price of Anarchy with network connectivity has a unimodal pattern, which reaches a peak median value at $M = 0.36$. As has also been shown previously, the level of dispersion of the Price of Anarchy across networks within each ensemble is positively correlated with higher values of the Price of Anarchy across the spectrum.

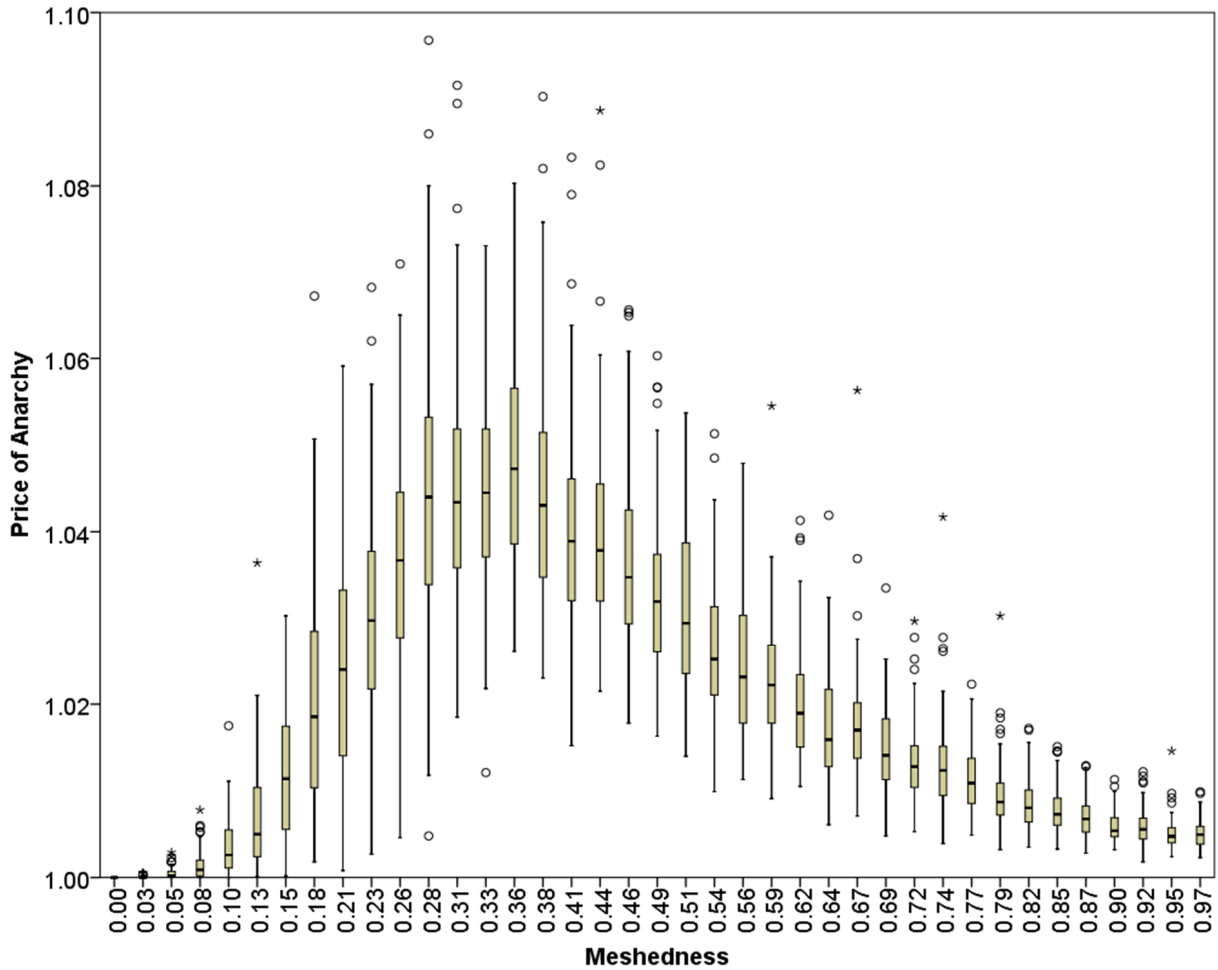

Figure 17 - Price of Anarchy against Meshedness $M$ ($A = 6.25$, $n = 100$, $\rho_n = 16$ and $\rho_{dem} = 2355$)

## 4.4 Discussion of Results

These results presented in the preceding section show that numerical investigations following the proposed experimental framework can provide substantial insights into how network structure affects performance. In particular, the results provide clearer relationships between aspects of network structure and performance indicators, which can motivate further research to understand the mechanisms that underpin the variations shown. This section provides an example of such analysis.

It can be seen that there are several similarities in the broad patterns exhibited by the two performance indicators in four experiments. For example, the first and second experiments both show that the average link V/C ratio increases monotonically with respect to increases in demand density and network size. The third and fourth experiments show that this measure also decreases monotonically with respect to increases in network density and network connectivity. Finally, all four experiments find a unimodal pattern for the Price of Anarchy. These similarities exist because all four experiments actually explore different ways of adjusting the balance between the total amount of travel demand and the total amount of network supply. In the first experiment, network supply remains unchanged as total demand increases. In the second experiment, total network supply and total travel demand increase together. In both the third and fourth experiments, total demand is fixed whilst network supply is increased.

With respect to the average link V/C ratio, it follows from these descriptions that the first, third and fourth experiments actually pick up on the same simple causal mechanism that governs the variation in this measure. This mechanism is that an increase in the ratio of total demand to total network supply leads to an increase in congestion, regardless of whether it is achieved by increasing demand or decreasing the amount of network supply, whilst a decrease in the ratio of total demand to total capacity leads to a decrease in congestion, regardless of whether it is achieved by decreasing demand or increasing capacity. This mechanism does not apply to the second experiment because total demand and total network supply increase at the same rate. The average link V/C ratio increases across this spectrum because, as network size increases, the total volume of flow using routes that pass through the geometric centre of the domain also increases. The volume of flow on links in the geometric centre of the domain therefore also increases, which creates congestion because the capacities of these links remain fixed as network size increases.

An explanation for the unimodal pattern in the Price of Anarchy, common to all four experiments, is less obvious. However, it does appear to be strongly connected to how the level of congestion changes across each spectrum as the ratio of total demand to total network supply changes. In particular, comparisons of the graphs for the two performance indicators in each experiment reveal that the peak regions of the Price of Anarchy coincide with values of the average link V/C ratio between approximately 0.5 and 0.8.

## 5  Conclusions

The aim of this paper was to explore how the approach of using large ensembles of synthetically generated networks could be better used to study how network structure affects network performance in urban road networks. An understanding of how different structures of network infrastructure and travel demand, combine to yield different performance characteristics would be useful for both transport policy and network design. In particular, such understanding could help identify how existing road networks in urban areas can be used more effectively (Mak and Rapoport, 2013), or how structural features, which yield desirable performance characteristics, could be built into the construction of new road networks.

Towards this aim, this paper has reviewed empirical studies of network structure, finding that road traffic networks have several small-scale similarities but that there is also considerable variation with respect to network size, density and connectivity. This paper has also reviewed studies of the effects of network structure on performance, finding that existing studies use structures of supply that are not plausible for urban road networks, and that the existing experimental approach does not provide a coherent picture of how network structure affects performance. In response to these limitations, this paper has proposed an experimental framework for the design of experiments to investigate how network structure affects performance.

Inspired by empirical studies, this paper has then demonstrated an application of this framework to study how two performance indicators; the average link V/C ratio and the Price of Anarchy, vary with respect to the density of travel demand, and the size, density and connectivity of network supply structure. To achieve this, this paper proposed a simple model of road network generation that is able to generate networks with a wide range of structural properties, including ranges of values that have been observed by empirical studies of real road networks.

In demonstrating the application of this framework, we have set out a more comprehensive and systematic approach to the study how road network structure affects performance. Unlike previous numerical studies in Network Science, this approach takes advantage of the findings of empirical studies of network structure to motivate specific research questions. It also enables the generalizability of findings to be evaluated because performance phenomena are connected to specific aspects of network structure.

There are several ways in which the example application of this experimental framework could be further developed. Firstly, the road network model could be refined so that it produces networks that reflect more of the properties of real road networks, such as those observed in empirical studies. The models of Barthelemy and Flammini (2009) and Courtat et al. (2011) provide useful starting points. A more sophisticated traffic model could also be used, to provide a more realistic representation of traffic flow and congestion e.g. to include junction delays. Any increase in the complexity of the model would necessarily increase the computational burden, which is already substantial. All of the experiments in this paper were conducted on a Windows PC and would benefit from a computing environment more suited to large scale calculation work, especially if more advanced models are to be used. Such a development would also enable bigger networks, perhaps on the scale of real road networks, to be tested. The key point is that the experimental framework is general enough to accommodate such changes.

This paper has noted several areas for future research. For example, existing empirical studies do not study how the structural properties of link capacities or junction types vary between urban areas. Such data would allow more realistic supply structures to be used in numerical experiments. We note that Open Street Map data offers a promising source of data for further work in this area; see, for example, Corcoran and Mooney (2013). There are also many other aspects of the effects of network structure on performance that this paper did not explore. For example, the effects of the distribution of travel demand on network performance remain largely unstudied. We also note that other performance indicators could be used, perhaps relating to other important aspects of performance such as environmental or safety indicators. Finally, we highlight that with appropriate changes to the modelling approach, the framework proposed in this paper could also be applied to studies of the effects of network structure on performance in other transport systems; for example, metros, buses and trains.